# GDCNet: Calibrationless geometric distortion correction of echo planar imaging data using deep learning

Marina Manso Jimeno[1,2], Keren Bachi[3,4], George Gardner[3,4], Yasmin L. Hurd[3,4], John Thomas Vaughan, Jr.[1,2,5,6], Sairam Geethanath[2,7]*

[1] Department of Biomedical Engineering, Columbia University in the City of New York, New York, NY, USA

[2] Columbia Magnetic Resonance Research Center, Columbia University in the City of New York, New York, NY, USA

[3] Department of Psychiatry, Icahn School of Medicine at Mount Sinai, New York, NY, USA

[4] Addiction Institute at Mount Sinai, Icahn School of Medicine at Mount Sinai, New York, NY, USA

[5] Department of Radiology, Columbia University Medical Center, New York, 10032, NY, USA

[6] Zuckerman Institute, Columbia University in the City of New York, New York, 10027, NY, USA

[5] Department of Radiology and Radiological Science, John Hopkins University, Baltimore, MD, USA

**Corresponding author:**
Sairam Geethanath
sairam.geethanath@jhu.edu
601 N Caroline St,
Baltimore, MD, 21205

## Conflict of interest statement

## Abstract

Functional magnetic resonance imaging techniques benefit from echo-planar imaging's fast image acquisition but are susceptible to inhomogeneities in the main magnetic field, resulting in geometric distortion and signal loss artifacts in the images. Traditional methods leverage a field map or voxel displacement map for distortion correction. However, voxel displacement map estimation requires additional sequence acquisitions, and the accuracy of the estimation influences correction performance. This work implements a novel approach called GDCNet, which estimates a geometric distortion map by non-linear registration to $T_1$-weighted anatomical images and applies it for distortion correction. GDCNet demonstrated fast distortion correction of functional images in retrospectively and prospectively acquired datasets. Among the compared models, the 2D self-supervised configuration resulted in a statistically significant improvement to normalized mutual information between distortion-corrected functional and $T_1$-weighted images compared to the benchmark methods FUGUE and TOPUP. Furthermore, GDCNet models achieved processing speeds 14 times faster than TOPUP in the prospective dataset.

# Abbreviations

**CNN**
Convolutional neural network

**DL**
Deep learning

**DWI**
Diffusion-weighted imaging

**EPI**
Echo-planar imaging

**FA**
Flip angle

**fMRI**
Functional magnetic resonance imaging

**GDM**
Geometric distortion map

**GE**
Gradient-echo

**ID**
In-distribution

**IRB**
Institutional Review Board

**MSE**
Mean squared error

**NMI**
Normalized mutual information

**OOD**
Out-of-distribution

**PE**
Phase encoding

**PSF**
Point spread function

**PSNR**
Peak signal-to-noise ratio

**SE**
Spin-echo

**ssEPI**
Single-shot echo-planar imaging

**SSIM**
Structural similarity index measure

**STU**
Spatial transform function unit

**TE**
Echo time

**TI**
Inversiton time

**TR**
Repetition time

**$T_1$w**
$T_1$-weighted

**VDM**
Voxel displacement map

## 1. Introduction

Functional magnetic resonance imaging (fMRI) has advanced our understanding of brain function by non-invasive mapping of neural activity[1,2]. Single-shot echo-planar imaging (ssEPI)[3] is widely employed in fMRI due to its rapid acquisition capabilities, making it suitable for capturing dynamic brain processes with high temporal resolution[4,5]. However, its low bandwidth along the phase encoding (PE) direction and prolonged read-out duration render it sensitive to inhomogeneities in the main magnetic field $B_0$, resulting in the accumulation of phase errors[6–9]. These errors lead to geometric distortion, which appears as shifts in local positions and stretched or compressed areas in the images along the PE direction, posing challenges to the spatial localization of brain activations[10,11]. Additionally, $B_0$ inhomogeneities cause shortening of $T_2^*$ relaxation, resulting in signal loss in long echo time (TE) gradient-echo (GE) sequences, commonly employed in fMRI. Techniques like multi-shot EPI[12,13] and point spread function (PSF)-EPI[14,15] include modifications to the k-space trajectory to mitigate these artifacts, albeit at the expense of acquisition time and temporal resolution. Consequently, distortion correction strategies are often favored and applied during data pre-processing[16–18].

Traditional distortion correction methods estimate a voxel displacement map (VDM), which depicts the shift of each voxel along the PE direction required to unwrap the distorted images. The VDM can be estimated using multiple approaches, the most common being from a field map acquisition[6,10], or a pair of reversed-PE-polarity EPI sets[19,20]. Field map-based methods estimate the $B_0$ field spatial distribution from dual-echo GE acquisitions. Equation 1 illustrates the computation of the VDM from the field map $\Delta B(x,y)$ and the bandwidth along the PE direction $BW_{PE}$, where $T$ is the echo spacing and $N_y$ is the number of phase encodes. Multi-echo

EPI[12,13] approaches estimate dynamic field maps from the EPI data. The dynamic correction is more robust to patient motion, but the estimated field maps are susceptible to distortion and other artifacts[21], impacting the correction performance. In contrast, the reversed encoding method involves acquiring a pair of EPI sets with opposite PE polarity, commonly dubbed "blip-up/blip-down" acquisitions. This method assumes that the distortions in these pairs of images have equal magnitude but opposite directions[19]. The VDM is estimated voxel-wise by calculating the displacement that minimizes the error between them. Existing tools compute it using discrete cosine basis functions[19] (TOPUP in the FSL toolbox[22]) or by solving a non-linear registration cost function[20]. Additionally, reversed encoding methods may include signal intensity correction and non-linear regularizers to anatomical reference[23–25]. Nevertheless, voxel-wise iterative optimization is time-consuming and computationally burdensome.

$$VDM = \frac{\Delta B(x,y)}{BW_{PE}} = \Delta B(x,y) \cdot T \cdot N_y \qquad (1)$$

VDM estimation from a field map or reversed-encoded EPI requires additional scans and relies on a static VDM to correct the fMRI time-series data, leading to prolonged scanning times and susceptibility to errors arising from motion-induced $B_0$ changes[26]. Another distortion correction approach involves non-linear registration to available non-distorted anatomical images[27,28], traditionally $T_1$-weighted ($T_1$w) images commonly acquired in fMRI imaging protocols[29–31] and less susceptible to $B_0$ perturbations. Although this method avoids additional acquisitions, it is computationally intensive and slow in practice[32,33].

Deep learning-based methods have been employed for EPI distortion correction. In supervised training, Liao et al.[34] and Hu et al.[35] proposed convolutional neural networks (CNN) to correct ssEPI distortions using simulated data and PSF-EPI images as ground truth, respectively. However, the dependence on ground truth data presents challenges as undistorted EPI images are difficult to obtain, and simulation-based results may not accurately reflect real clinical scenarios. Since the introduction of the unsupervised medical image registration framework VoxelMorph[36], several studies have explored its potential for correcting EPI susceptibility artifacts. VoxelMorph's architecture involves a U-Net[37] that predicts a displacement field based on two input volumes and a spatial transform function that unwraps the moving input volume to align it with the fixed target volume. Leveraging a similar approach, S-Net[38] addressed susceptibility artifacts in 3D reversed-PE EPI images. It applies the predicted displacement field with opposite signs to unwrap the two input volumes and adds a similarity loss function between the outputs. Legouhy et al.[39] proposed a semi-supervised method incorporating Jacobian intensity modulation and a supervised loss for the displacement map. Deep flow net[40] added density weighting to the unwrapping operation to jointly correct signal intensity errors. Additionally, it computed the smoothing loss function at two image resolutions, including a penalty term to restrict the displacement map within reasonable ranges. FD-Net[41] and DLRPG-net[42] both leverage a forward-distortion module and have an additional similarity loss term calculated against the inputs. Moreover, DLRPG-net predicts the displacement map by estimating the coefficients of cubic spline vectors and includes a loss term that maximizes normalized mutual information (NMI) against an anatomical image. However, all studies[38–42] required the acquisition of reversed-encoded EPI images and used Spin-Echo (SE) EPI images as inputs. Reversed-PE SE-EPI acquisitions have a direct application for distortion correction of diffusion-weighted or diffusion tensor imaging (DWI) or (DTI) but their application to correction of GE-EPI fMRI images has not been tested with this approach.

We present a technique partly motivated by VoxelMorph[36], for dynamic distortion correction of GE-EPI images. Our method estimates a geometric distortion map (GDM) from ssEPI fMRI and $T_1$w images and apply it for distortion correction of the functional data. By utilizing the readily available $T_1$w anatomical data, we avoid the need for additional sequence acquisitions, reducing scan time and improving efficiency.

## 2. Methods

### 2.1. Network architecture and implementation

Let $EPI_{GD}$ and $T_1w$ be the two input $n$D images with $n = 2, 3$. We leverage a CNN to map the function $f_\theta(T_1w, EPI_{GD}) = GDM$, where $\theta$ are the network parameters. The input images are and they are concatenated as a two-channel magnitude image with dimensions 64x64x2 or 64x64x32x2 for 2D and 3D images, respectively. The GDM is an $n$D image that reflects the displacement field that registers the input images along the PE direction (first dimension $i$).

The CNN architecture follows a U-Net, consisting of an encoder and a decoder path with skip connections between the two, as depicted in Figure 1. We apply 2D or 3D convolutions across all layers of the CNN depending on the input dimensions with a kernel size of 3, followed by LeakyReLU activations. We add a stride of 2 to the convolutions of the encoding layers to reduce the image dimensions by half, whereas, in the decoding path, we include upsampling blocks between convolution layers to obtain an output GDM of the same spatial resolution as the input images. The outputs of the decoding and encoding layers with equal image resolution are concatenated via skip connections to reduce the effects of vanishing gradients during training and retain high-frequency information. This configuration allows the localization of the distortion regions and enables correction of the EPI images.

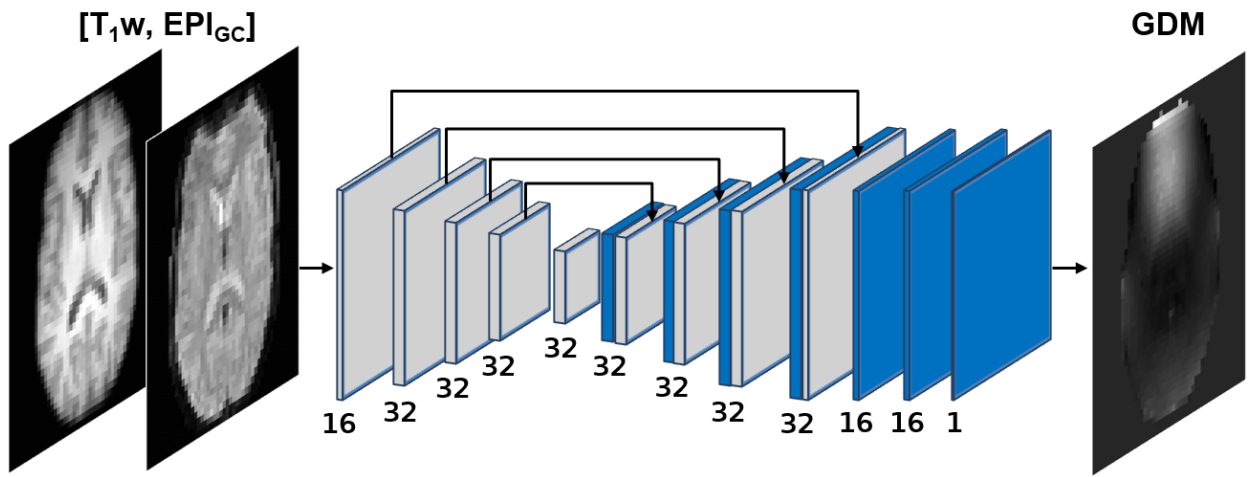

**Figure 1.** U-Net architecture. The two inputs are concatenated before the first convolutional layer. The number below each layer corresponds to the number of filters, and the arrows represent skip connections and concatenation between the encoder and decoder layers. The last layer is a single-filter convolution to output the GDM along the phase encoding dimension. GDM: geometric distortion map.

The output of the U-Net and the EPI image are then fed to the unwrapping module to generate the distortion-corrected image $EPI_{GDC}$. The final objective is to minimize the cost function between $EPI_{GDC}$ and $T_1w$. The network parameters are updated after computing the loss; therefore, the unwrapping module operations must be differentiable. Based on the GDM, the unwrapping module (Figure 2) computes the voxel location shift and applies it to the distorted image. It approximates the voxel values in the unwrapped image by linear interpolation of the values of neighboring voxels at the original voxel position. Since EPI distortions are negligible along the frequency encoding direction, the STU only unwraps the image along the PE direction.

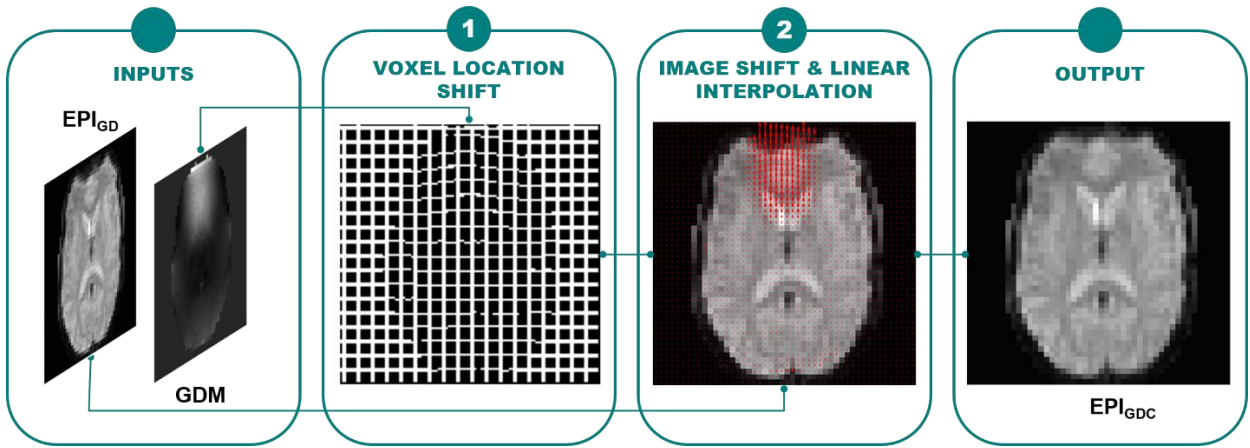

**Figure 2.** The unwrapping module applies the spatial transformation or image shift (red arrows) provided by the input GDM to the distorted EPI image. After transformation, linear interpolation approximates the voxel intensities in the distortion-corrected image. EPI: echo-planar imaging; GDM: geometric distortion map.

We trained the network in supervised, semi-supervised, and self-supervised formulations. Inputs and outputs for each of the model implementations and loss functions are depicted in Figure 3. The supervised network consists exclusively of the U-Net, and its loss function minimizes the mean squared error (MSE) voxel-wise difference between the estimated GDM and the ground truth VDM. The semi-supervised model includes the differentiable STU and adds a self-supervised term to the loss function. This loss term penalizes differences in appearance between the distortion-corrected and anatomical reference $T_1$w images by optimizing their local cross-correlation[43]. We utilized this loss function because of its robustness to intensity variations due to the difference in contrast between $T_1$w and $T_2^*$-weighted GE-EPI images. Finally, the self-supervised model substitutes the supervised MSE loss term with a smoothness constraint that penalizes abrupt spatial variations in the estimated GDM using a diffusion regularizer on its spatial gradients[36].

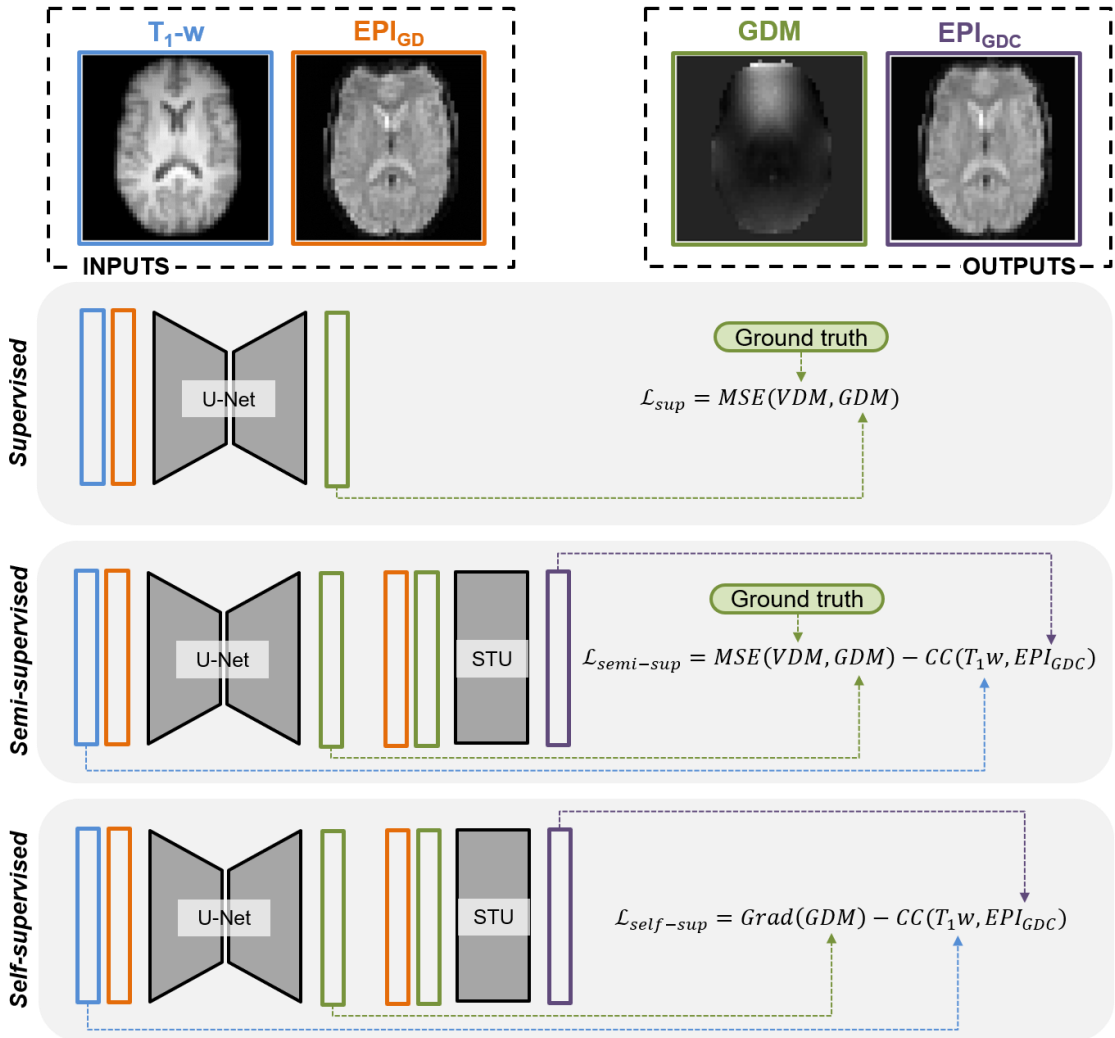

**Figure 3.** Network diagrams and loss functions for the supervised, semi-supervised, and self-supervised implementations. Inputs and outputs are color-coded with the legend at the top of the figure. The supervised and semi-supervised configurations minimize MSE between the estimated GDM and ground truth VDM. The semi-supervised and self-supervised models include a loss term that penalizes dissimilarity between the corrected EPI and anatomical reference image. Additionally, the self-supervised loss substitutes the supervised loss term by a smoothness constraint that penalizes abrupt spatial variations in the estimated GDM. EPI: echo-planar imaging; GDM: geometric distortion map; MSE: mean squared error; VDM: voxel displacement map.

The three model configurations were trained on 2D and 3D data, rendering six models in total. Due to the large number of models, we primarily report the results of the self-supervised models in this work, as they achieved the best distortion correction performance and do not require ground truth data for training. The results of the other models are reported in the Supplementary Material. We implemented our method using the Tensorflow-Keras[44] framework and ADAM optimizer[45] with a learning rate of $10^{-5}$. We trained our models for 100 epochs with a batch size of 1. The code for our implementation is available in GitHub[46].

## 2.2. Datasets

### A) Retrospective

We used four publicly available OpenNeuro datasets to train and test our models. The total number of subjects across datasets is 48, each including at least a reference $T_1$w anatomical scan, a GE-EPI resting-state fMRI session, and magnitude and phase difference images to calculate a field map. Certain acquisition parameters, subject age ranges, and health conditions are different for each dataset. The most relevant parameters for each dataset are summarized in Table 1. All EPI images had an in-plane matrix size of 64x64. Image pre-processing was performed using the FSL toolbox and included brain extraction and co-registration of $T_1$w images to the EPI data space, field map calculation, selection of ten time-frames out of the EPI functional data, and their geometric distortion correction using FUGUE. Before training, $T_1$w and EPI images were scaled to 32 brain slices and normalized to the range [0, 1], and the ground truth VDM was computed from the field map as in Equation 1. All ten subjects from ds000224 were considered only for out-of-distribution testing. We split the remaining 38 subjects' data 80%-20% for training and in-distribution testing, respectively.

**Table 1.** Demographics and acquisition parameters for the retrospective dataset.

| Dataset Number | Number of Subjects | Age [years] | Male/Female [%] | Condition | Scanner | Isotropic resolution [mm] | $BW_{PE}$ [Hz] |
|---|---|---|---|---|---|---|---|
| 1: ds000224[47] | 10 | 24-34 | 50/50 | Healthy | Siemens Trio | 4 | 26.48 |
| 2: ds001454[48] | 24 | 19-38 | 42/58 | Healthy | Siemens Skyra | 3 | 47.35 |
| 3: ds002799[49] | 9 | 30-50 | 44/56 | Refractory epilepsy | Siemens Trio | 3 | 27.9 |
| 4: ds004101[50] | 5 | 43-61 | 0/100 | Emotional dysregulation | Siemens Verio | 3 | 30.64 |

*B) Prospective*

We tested our models on a prospective dataset of 2 sessions from 15 subjects acquired in a 3 T Siemens MAGNTETOM Skyra. Before data acquisition, all subjects provided written informed consent per the protocol approved by the Institutional review board (IRB). This data included T$_1$w data (TE=2.07 ms, TR=2.4 s, TI=1 s, isotropic resolution=0.8 mm), two sets of SE-EPI acquisitions (TE=65 ms, TR=8.6 s, FA=80°, isotropic resolution=2.1 mm) with reversed PE polarity and a resting state and task-based GE-EPI acquisition (TE=35 ms, TR=1 s, FA=60°, isotropic resolution=2.1 mm) with at least 600 time-points (approximately 10 minutes of acquisition time). The bandwidth along the PE direction for the EPI acquisitions was 13.62 Hz. A total number of 50 2D GE-EPI fMRI acquisitions across subjects and scans were considered for correction using GDCNet. We calculated the field map and ground truth VDM from the reversed-encoded data using FSL's TOPUP and used it as a benchmark method. All images were downsampled from 108x108 to 64x64 to fit the model's input dimensions prior to inference. The remaining pre-processing steps were the same as for the retrospective datasets.

## 2.3. Experiments and evaluation metrics

We conducted a regularization analysis on the self-supervised models to refine the weight of the smoothness loss term. We investigated its impact on the distortion correction performance by training the model for 25 epochs with various values of the regularization parameter $\lambda$.

The performance evaluation of the proposed GDCNet models involved qualitative and quantitative assessments using three different testing scenarios: in-distribution (ID) testing, out-of-distribution (OOD) testing, and prospective testing. For ID testing, we used a subset of 20% of the subjects from datasets 2-4 (Table 1). In contrast, the OOD test set included all ten subjects from dataset 1. This experiment evaluated the generalization capability of the models on data that differed from the training distribution. Prospective testing was carried out to assess the real-world performance of the model on new, unseen data acquired prospectively and generalization to differences in acquisition parameters.

To assess EPI distortion correction performance, we compared GDCNet unwrapped EPI images with the corrected images obtained by the benchmark methods, namely FUGUE and TOPUP. FUGUE was utilized when field maps were available, whereas TOPUP was used with reversed-PE data. We calculated SSIM and PSNR for quantitative analysis. Additionally, we computed NMI to evaluate the anatomical alignment of the EPI images to the $T_1$w image before and after correction with GDCNet and the benchmark methods. NMI values range from 0 (no mutual information) to 1 (complete correlation). All metrics were computed slice-wise. Additionally, we compared the processing speeds of GDCNet and TOPUP on the prospective dataset. This experiment was performed on an Intel Xeon ES-2690 v4 CPU with one NVIDIA Quadro M6000 24 GB GPU.

Statistical analysis to compare distortion correction performance between the GDCNet models and the benchmark methods included a one-way ANOVA test of the mean NMI among the groups, followed by post-hoc analysis using Tukey's test to determine specific group differences. The null hypothesis for the ANOVA test was that there were no significant differences in the mean NMI among the correction methods. Subsequently, the null hypothesis for each pairwise comparison was that there was no significant difference in the mean NMI between the two groups being compared.

## 3. Results

### 3.1. Regularization analysis

We evaluated the self-supervised models using varying values of the smoothness regularization parameter $\lambda$ on the ID and OOD test sets. **¡Error! No se encuentra el origen de la referencia.** illustrates the resulting average Dice scores for each model, along with a representative slice for their predicted GDM and distortion corrected EPI images. Not imposing any constraint on the GDM ($\lambda=0$) resulted in large and unrealistic displacement of signal in the corrected images. Among the remaining models, both quantitative and qualitative analyses indicated that setting $\lambda=0.5$ as regularization parameter yielded optimal distortion correction. However, the reduced smoothness of the GDM caused voxel displacement in undistorted regions of the brain, leading occasionally to undesired changes in brain anatomy, such as a shrinkage of the ventricles size in the example in Supplementary Figure 1. We opted for a more conservative approach and the results presented here correspond to models trained with $\lambda=1$. Training for 100 epochs took 4 h 6 min and 39 min for the 2D and 3D self-supervised models, respectively.

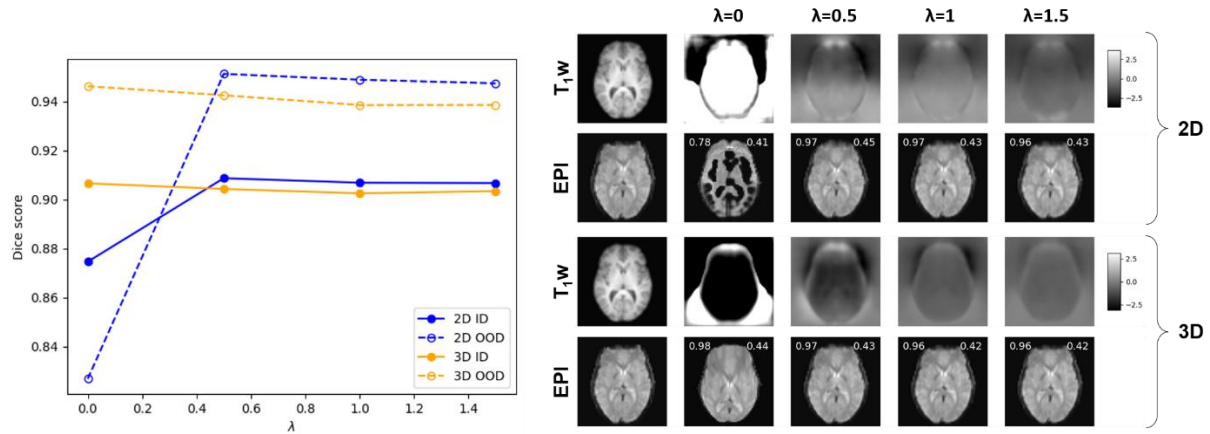

**Figure 4.** Quantitative and qualitative results of the regularization analysis of the 2D and 3D self-supervised models evaluated on the in-distribution and out-of-distribution test sets. For visualization purposes, we chose the display range of the GDM based on the absolute maximum displacement of the predicted GDMs of the models trained with $\lambda$=0.5, 1, and 1.5.

## 3.2. EPI distortion correction

The unwrapping of EPI images is achieved by the STU integrated within the semi-supervised and self-supervised models, or performed independently as a post-processing step utilizing the output of the U-Net in the supervised models. Representative slices of the unwrapping results from the ID and OOD test sets are shown in Figure 5. Upon visual inspection, the unwrapped images demonstrate better alignment to $T_1$w images. Areas of voxel stretching, pointed out by the arrows in Figure 5, are notably improved by the self-supervised models, whereas for areas of voxel compression (rectangles in Figure 5), the correction is analogous to FUGUE. Supplementary Figure 2 shows the distortion correction results from the supervised and semi-supervised models.

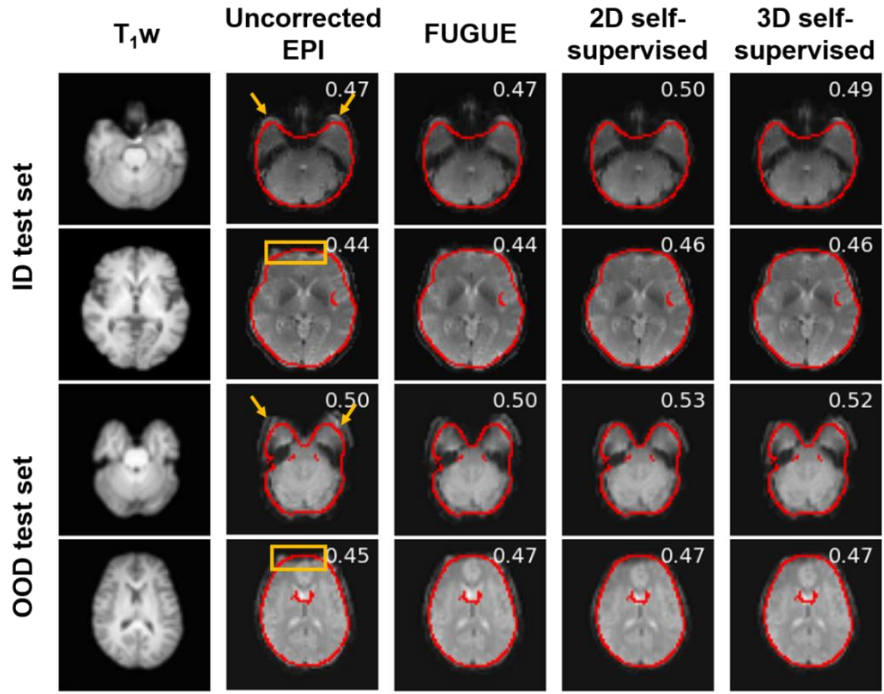

**Figure 5.** EPI distortion correction results for representative slices of the in-distribution and out-of-distribution test sets. The top-right numbers over the images are NMI values computed with respect to the T$_1$-weighted images. The red contours correspond to the edges of the T$_1$-weighted images, overlaid over the models' results for visual reference. The colored arrows and rectangles indicate areas of distortion, such as voxel stretching, compression, and signal loss. EPI: echo-planar imaging; NMI: normalized mutual information.

Quantitative results revealed that the self-supervised models achieve unwrapped images with improved NMI compared to uncorrected EPI images and corrections using the reference method FUGUE for both the ID and OOD test sets (Figure 6). The unwrapped images achieved average SSIM values above 0.85 and PSNR values of approximately 25 dB compared to FUGUE corrections. Supplementary Figure 3 displays the quantitative analysis of distortion correction results for all models. The corrections of the supervised models show modestly higher NMI than the uncorrected images but are outperformed by FUGUE. In contrast, the semi-supervised models achieved NMI values analogous to FUGUE.

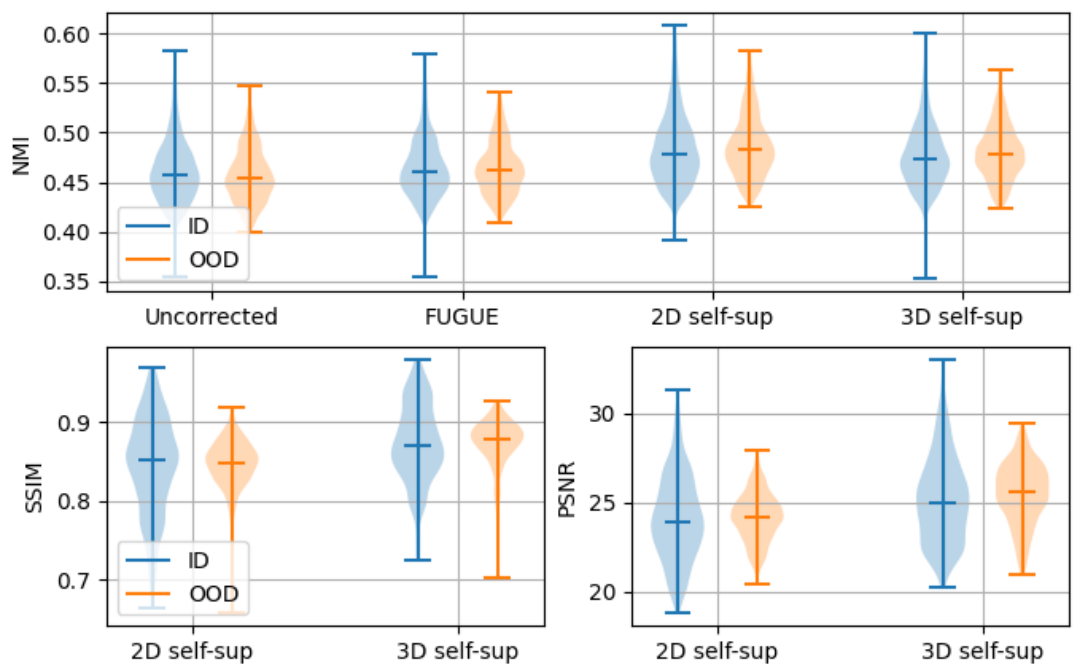

**Figure 6. (Top)** NMI values of the EPI and $T_1$-weighted images before and after correction with FUGUE and the models' correction results on the in-distribution and out-of-distribution tests. **(Bottom)**: SSIM and PSNR values of the models' EPI geometric distortion correction results on the in-distribution and out-of-distribution test sets with respect to the benchmark FUGUE correction. NMI: normalized mutual information; EPI: echo-planar imaging; SSIM: structural similarity index measure; PSNR: peak signal-to-noise ratio.

### 3.3. Prospective testing

Figure 7 illustrates the results of the qualitative analyses on the prospective dataset. Visual examination of the GDM predicted by the self-supervised models indicate increased similarity to the TOPUP-estimated VDM than those derived from field map acquisitions. Both methods achieved similar distortion correction performance. The predicted GDMs are smoother, and lack regions of abrupt magnitude changes within the brain, such as regions of large tissue susceptibility differences. TOPUP correction resulted in overstretching in areas where EPI images exhibit signal loss artifacts (yellow circle in Figure 7). Such effect was avoided by the self-supervised models' corrections. Supplementary Figure 4 includes inference results for the other four models.

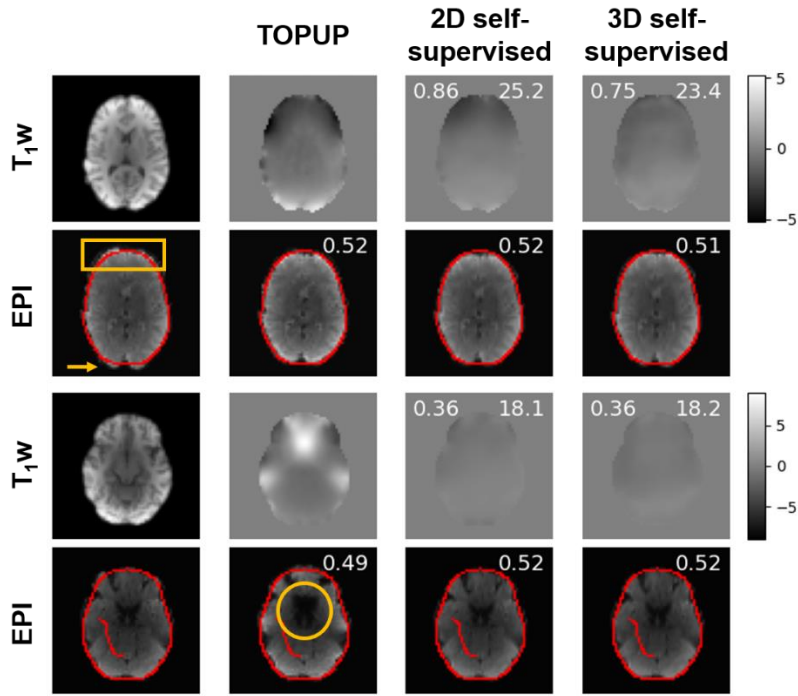

**Figure 7.** GDM prediction and EPI geometric distortion correction results of representative slices of the prospective test set. The top-left and top-right numbers over the estimated GDMs correspond to SSIM and PSNR values computed using TOPUP estimated VDM as reference image. The top-right numbers over the EPI-corrected images are NMI values computed with respect to the $T_1$-weighted images. The red contours correspond to the edges of the $T_1$-weighted images, overlaid over the models' results for visual reference. The colored arrow, rectangle, and circle indicate areas of voxel stretching, compression, and signal loss, respectively. GDM: geometric distortion map; EPI: echo-planar imaging; VDM: voxel displacement map; SSIM: structural similarity index measure; PSNR: peak signal-to-noise ratio; NMI: normalized mutual information.

Quantitatively, mean SSIM and PSNR values exhibited improvement for GDM prediction on the prospective set (Figure 8) compared to the ID and OOD test sets. Regarding distortion correction, NMI values indicated that both self-supervised models outperformed all the other models (Supplementary Figure 5) and the benchmark method for the prospective test set TOPUP.

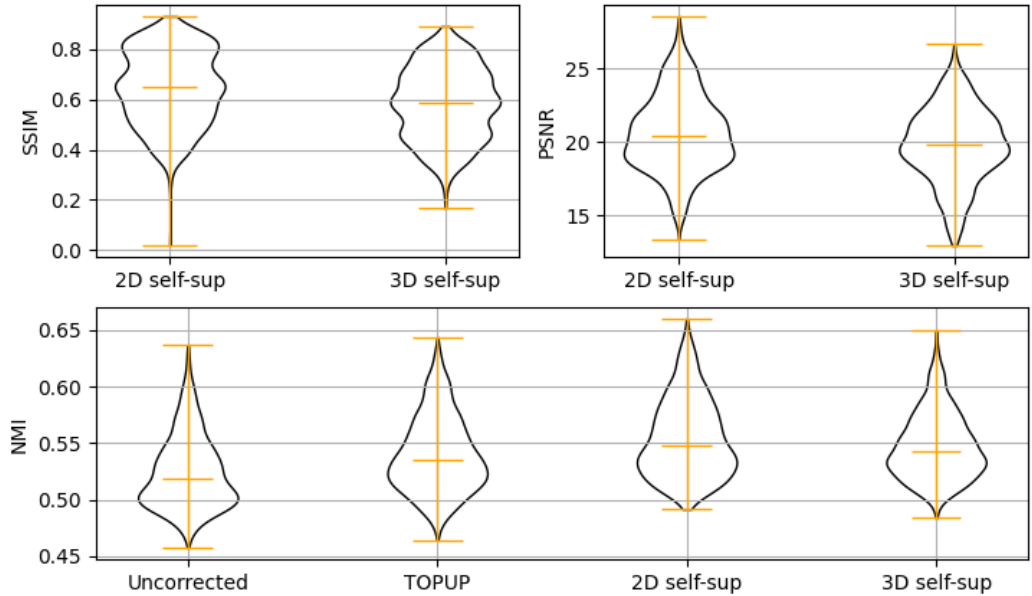

**Figure 8: (Top)** GDM prediction and **(Bottom)** EPI geometric distortion correction results for the prospective test set. We calculated NMI between the correction methods and the $T_1$w images. SSIM and PSNR metrics used the benchmark TOPUP method as reference images. GDM: geometric distortion map; EPI: echo-planar imaging; NMI: normalized mutual information; SSIM: structural similarity index measure; PSNR: peak signal-to-noise ratio.

Table 2 presents the runtime results of the proposed models and TOPUP for the correction of the prospective dataset. Supplementary Table 1 includes the runtime results of all six models. The benchmark method includes reversed-PE SE-EPI acquisition, VDM estimation, and unwrapping of the images using the resulting estimated field map. The average processing time was 14:04 (minutes:seconds) for an EPI acquisition of 600 time-points using a standard implementation of FSL running in the CPU. In contrast, all GDCNet models provided faster processing times on the same dataset and eliminated the need for additional sequence acquisition for GDM estimation. The 2D supervised model had the slowest processing speed, taking approximately 8 minutes, while the 3D semi- and self-supervised models offered the fastest processing time, completing VDM estimation and EPI distortion correction in 13 seconds.

**Table 2.** Acquisition, computation time, and processing speed for sequence acquisition, VDM or GDM estimation, and EPI correction.

|                              | TOPUP          | 2D self-supervised | 3D self-supervised |
|------------------------------|----------------|--------------------|--------------------|
| Sequence acquisition [s]     | 34             |                    |                    |
| VDM/GDM estimation [s]       | 432.53 ± 2.13  | 74.27 ± 1.56       | 12.63 ± 0.08       |
| EPI correction [s]           | 377 ± 6.18     |                    |                    |
| Total time [minutes:seconds] | 14:04          | 01:14              | 00:13              |

Processing times are reported as mean ± standard deviation of 15 fMRI datasets of 600 time-points each. Note that for TOPUP, a single VDM is used to correct all the time frames of the functional data. In contrast, the proposed models estimate a GDM for each time frame.

### 3.4. Statistical analysis

Table 3 presents the results of the statistical comparison of the self-supervised GDCNet models versus the images before and after correction using FUGUE or TOPUP baseline methods on the three test sets. Supplementary Table 2 includes the statistical analysis of all six models. The tests with p-values below 0.05 are highlighted in bold, and for such cases, we rejected the null hypothesis with 95% confidence. These p-values were adjusted for false discovery control using the Benjamini-Hochberg method. Both models demonstrated significant improvements to NMI compared to uncorrected data and compared to FUGUE and TOPUP in all test sets.

**Table 3.** Statistical analysis of mean NMI between the GDCNet models, uncorrected images, and benchmark corrections.

|                     |                            | 2D self-supervised |         | 3D self-supervised |         |
|---------------------|----------------------------|--------------------|---------|--------------------|---------|
|                     |                            | Mean diff.         | Sig.    | Mean diff.         | Sig.    |
| ID test set         | *Uncorrected* vs. GDCNet   | 0.02               | **0.005** | 0.014            | **0.005** |
|                     | *FUGUE* vs. GDCNet         | 0.017              | **0.005** | 0.011            | **0.009** |
| OOD test set        | *Uncorrected* vs. GDCNet   | 0.03               | **0.001** | 0.023            | **0.001** |
|                     | *FUGUE* vs. GDCNet         | 0.022              | **0.001** | 0.015            | **0.001** |
| Prospective test set| *Uncorrected* vs. GDCNet   | 0.03               | **0.001** | 0.023            | **0.001** |
|                     | *FUGUE* vs. GDCNet         | 0.014              | **0.001** | 0.007            | **0.001** |

A p-value below 0.05 (bold) indicates that the null hypothesis of equal mean NMI is rejected with 95% confidence level. The p-values in red indicate the models for which the mean NMI improved compared to no correction or correction with a benchmark method with statistical significance.

## 4. Discussion

In this study, we evaluated the performance of the proposed GDCNet models for distortion correction of GE-EPI-based fMRI images by non-linear registration to a non-distorted anatomical reference image. Our approach explored and compared six neural networks that estimate a GDM from the distorted EPI and $T_1$w images in a supervised, semi-supervised, and self-supervised manner, depending on the loss term minimized during training. We trained three models slice-wise or 2D and three models volume-wise or 3D. Since we used the same training data for both types of models, the training of the 3D models took less time, as 32 slices were processed in a single batch. The training and testing data included EPI data of healthy and pathological subjects acquired in multi-site, multi-model 3 T Siemens systems with varying ranges in PE bandwidth (Table 1). Our models demonstrated the ability to mitigate geometric distortion without additional sequences, such as dual-echo GE or a pair of reversed-PE EPI acquisitions. Instead, the GDCNet models leverage $T_1$w images, which are always included in neuroimaging studies for structural analysis and anatomical reference[29–31,51]. This feature and the contrast-independent nature of the cross-correlation loss function used to train the semi and self-supervised models make this approach adaptable to the correction of other EPI-acquired images, such as those for DWI and DTI.

VDM hotspots indicate areas where distortions are more pronounced, such as regions near air-tissue interfaces. When estimated from reversed-encoded EPI or dual-echo GE images, pixel distortions in the VDM are proportional to perturbations to the main magnetic field $B_0$. As the self-supervised models estimate the GDM by non-linear registration of the input images, the prediction of these models reflects only regions of distortion with respect to the $T_1$w images and does not reflect all hotspots of $B_0$ perturbations. Therefore, it cannot be considered proportional to the $B_0$ map as introduced in Equation 1. The predicted GDMs were more similar to TOPUP-estimated VDMs compared to dual-echo GE field maps (Supplementary Figure 6). The increase in SSIM and PSNR may stem from the fact that both types of displacement maps were estimated only along the PE direction. Additionally, these maps are not masked to remove background regions as dual-echo GE field maps do, improving distortion correction in cases where the EPI images stretch out of the anatomy boundaries of the $T_1$w images.

The self-supervised models exhibit the best distortion correction performance across the three test sets measured by the alignment between the EPI corrected and the anatomical $T_1$w images. These models achieved statistically significant improvements in NMI compared to traditional methods such as FUGUE and TOPUP (Table 3). In particular, the self-supervised models perform best at correcting areas of voxel stretching where the EPI voxels stretch out of the contours of the $T_1$w images (arrows in Figure 5), and they avoid overstretching in regions of signal voids (circle in Figure 7) characteristic FUGUE and TOPUP correction. The benefit of self-supervised training compared to supervised or semi-supervised training and methods that rely on a field map is that their performance does not depend on the accuracy of the ground truth field map acquisition, which often may be impacted by phase unwrapping, signal loss, poor generalization to different acquisition parameters, poor alignment between both image volumes, and other errors[34,42,52].

Testing the models' performance on data unseen during training is essential to assess their ability to generalize to diverse data distributions and gauge their success in real-world deployment scenarios, where data is often heterogeneous. This study evaluated the proposed models on an OOD retrospective dataset and a prospective dataset acquired with different protocol and acquisition parameters. Figure 7 and Figure 8 demonstrate that the self-supervised models achieved the best EPI distortion correction compared to the other models, with statistically significant improvements in anatomical NMI over TOPUP correction (Table 3). This experiment demonstrates successful distortion correction of EPI images acquired with a very low PE bandwidth of 13.62 Hz. Although with non-statistically significant differences for the ID test set, the 2D models outperformed the 3D models in all cases. Our hypotheses are 1) the more extensive data size used to train the 2D models and 2) the effect of the smoothness constraint in the 3D self-supervised model. Slice-wise 2D models were trained using a dataset 32 times larger than the 3D training dataset, as each time point had 32 slices. Furthermore, the ground truth field maps were estimated from 2D acquisitions with a gap between slices, making 2D kernels more suitable for the VDM estimation task.

The runtime analysis on the prospective dataset demonstrated that all models outperform TOPUP in terms of processing time and savings to scan times due to the elimination of supplemental sequences for VDM estimation. The 3D semi- and self-supervised models exhibited the fastest processing time, completing the correction with an average of 13 seconds for an EPI acquisition of 600 time-points. Despite the faster computation of 2D convolutions, the 3D models show reduced inference time due to the ability to process a full time-point simultaneously instead of individual slices. The 2D self-supervised model completed GDM estimation and EPI correction of the same dataset in 1 minute and 15 seconds, approximately 14 times faster than TOPUP (Table 2). In all cases, the GDCNet models offer the advantage of dynamic correction of the functional series, whereas traditional methods correct it by leveraging a static VDM. Theoretically, dynamic VDM estimation renders a more robust EPI correction due to the reduced sensitivity to $B_0$ temporal changes and intra- and inter-sequence motion. Consequently, we expected that the standard deviation of NMI across time-points would be smaller after correction with the 2D self-supervised GDCNet model compared to TOPUP. This observation holds true for some prospective functional datasets (Figure 9), but we could not isolate these differences exclusively to motion robustness.

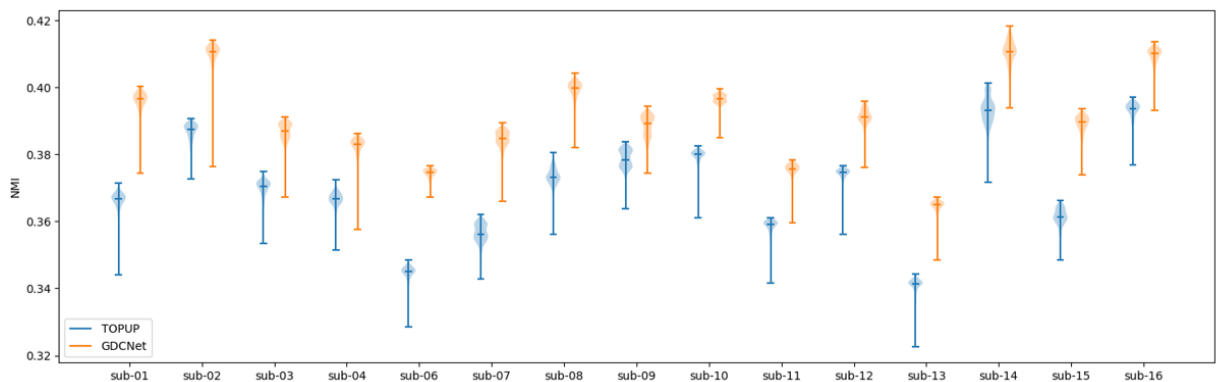

**Figure 9.** Volume-wise NMI distribution computed across the 600 time-points of 15 prospective functional datasets for the 2D self-supervised GDCNet model and TOPUP corrections. In all cases, GDCNet renders higher NMI values between the EPI corrected and the $T_1$-weighted reference images. NMI: normalized mutual information.

A limitation of this study is the fixed input size of 64x64 for the 2D models and 64x64x32 for the 3D models. Images with different acquisition matrices require resampling prior to inference. However, our experiment testing the models on the prospective dataset of input size 108x108 did not affect the performance of the self-supervised models. The other limitation is the dependence on the $T_1$w images, for which acquisition and the results of post-processing steps such as brain extraction and co-registration must be carefully evaluated and sometimes fine-tuned for each subject to avoid artifacts and downstream errors. Future studies will further investigate the differences between the 2D and 3D self-supervised models and investigate the potential robustness of GDCNet to motion.

## 5. Conclusion

GDCNet demonstrated fast EPI distortion correction of fMRI images without acquiring additional sequences for $B_0$ map estimation by non-linear registration to reference anatomical images. Among the compared models, the 2D self-supervised configuration resulted in the best NMI between distortion-corrected EPI and anatomical reference $T_1$w images, outperforming the benchmark methods FUGUE and TOPUP in both correction performance and processing speed.